\documentclass[journal,onecolumn]{IEEEtran}

\IEEEoverridecommandlockouts

\usepackage{amsmath,amssymb,amsfonts,amsthm}
\usepackage{graphicx}
\usepackage{multicol}
\newtheorem{thm}{Theorem}
\newtheorem{prop}{Proposition}
\newtheorem{lem}{Lemma}

\newtheorem*{WH}{Working hypothesis}
\newtheorem*{nota}{Notation}

\def\0{{\mathbf 0}}

\newcommand{\R}{\mathbb{R}}
\newcommand{\F}{\mathbb{F}}
\newcommand{\Z}{\mathbb{Z}}

\newcommand{\defeq}{\overset{\text{def}}{=}}

\begin{document}

\sloppy

%% Paper Title
%% You can use linebreaks \\ within to get better formatting as
%% desired. 
\title{Lattice Codes for  the Binary Deletion Channel}

%% Author names and affiliations:
%%
%% Avoiding spaces at the end of the author lines is not a problem with
%% conference papers because we don't use \thanks or \IEEEmembership.
%%
%% For several authors with only one affiliation:
%%
% \author{
%   \IEEEauthorblockN{Hui-Ting Chang and Stefan M.~Moser}
%   \IEEEauthorblockA{Department of Electrical and Computer Engineering\\
%     National Chiao Tung University (NCTU)\\
%     Hsinchu, Taiwan\\
%     Email: \{email-of-hui-ting,email-of-stefan\}@ieee.org} 
% }
%%
%% For up to three affiliations:
%%
\author
{
{Lin Sok$^{1}$, Patrick Sol\'e$^{1,2}$,  Aslan Tchamkerten$^{1}$\thanks{This work was supported in part by an Excellence Chair Grant
from the French National Research Agency (ACE
project).}}\\
%\author{Anonymous submission to DSD}

\small
$ ^1$Telecom ParisTech\\
$^2$ King Abdulaziz University\\
{\tt \{lin.sok;patrick.sole;aslan.tchamkerten\}@telecom-paristech.fr}\\

}

\date{}
%%
%% For over three affiliations, or if they all won't fit within the width
%% of the page, use this alternative format:
%%
% \author{
%   \IEEEauthorblockN{
%     Michael Shell\IEEEauthorrefmark{1},
%     Homer Simpson\IEEEauthorrefmark{2},
%     James Kirk\IEEEauthorrefmark{3}, 
%     Montgomery Scott\IEEEauthorrefmark{3} and
%     Eldon Tyrell\IEEEauthorrefmark{4}}
%   \IEEEauthorblockA{
%     \IEEEauthorrefmark{1}School of Electrical and Computer Engineering\\
%     Georgia Institute of Technology, Atlanta, Georgia 30332--0250\\ 
%     Email: see http://www.michaelshell.org/contact.html}
%   \IEEEauthorblockA{
%     \IEEEauthorrefmark{2}Twentieth Century Fox, Springfield, USA\\
%     Email: homer@thesimpsons.com}
%   \IEEEauthorblockA{
%     \IEEEauthorrefmark{3}Starfleet Academy, San Francisco, California 96678-2391\\
%     Telephone: (800) 555--1212, Fax: (888) 555--1212}
%   \IEEEauthorblockA{
%     \IEEEauthorrefmark{4}Tyrell Inc., 123 Replicant Street, Los Angeles, California 90210--4321}
% }

%% Use for special paper notices

%% To balance the two columns, you should reduce the text-height of
%% the last page using the following command:
%%%%%%%%%%%%%%%%%%%%%%%%%%%%%%%%%%%%%%%%%%%%%%%%%%%%%%%%%%%%%%%%%%%%%
%%\addtolength{\textheight}{-9.35cm}
%%%%%%%%%%%%%%%%%%%%%%%%%%%%%%%%%%%%%%%%%%%%%%%%%%%%%%%%%%%%%%%%%%%%%
%% with an appropriate value. This command must be place on the second
%% last page, i.e., for a one-page abstract here, for a two-page
%% abstract right after the \maketitle command.

%% Create the title:
\maketitle

%% Abstract: 
%% For the final version of the accepted paper, please make sure you
%% remove the comment "THIS PAPER IS ELIGIBLE FOR THE STUDENT PAPER
%% AWARD."
%%
\begin{abstract}
The construction of deletion codes for the Levenshtein metric is reduced to the construction of codes over the integers for the Manhattan metric by run length coding. 
The latter codes are constructed by expurgation of translates of lattices.
These lattices, in turn, are obtained from Construction~A applied to binary codes and $\Z_4-$codes. A lower bound on the size of our codes for the Manhattan distance are obtained through generalized theta series of the corresponding lattices.

\end{abstract}

%\maketitle
\section{Introduction}

Coding for the binary deletion channel remains a major challenge for coding theorists. Part of the reason for this is that the use of standard block algebraic coding techniques (parity-checks, cosets, syndromes) is precluded due to the specificity of the channel which produces output vectors of variable lengths. 
A variation of this channel is the so-called segmented deletion channel where at most a fixed number of errors can occur within segments of given size \cite{LM,MT}. Because of this restriction, the segmented deletion channel does not alterate the number of runlengths if they are long enough.  Hence, if we view the channel in terms of input/output runlengths, the input and output vectors have the same dimension (assuming long enough runlengths). In this case,  algebraic coding techniques can be used. 

In this paper, we construct lattice-based codes, which, in principle, can be decoded when obtained  via Construction A from Lee metric codes with known decoding algorithms \cite{CJC}. The proposed code constructions are analogous to the so-called  $(d,k)-$codes in magnetic recording where each codeword contains runs of zeros of length at least $d$ and at most $k$ while each run of ones has unit length \cite{LH}. Given $d,k$ 
and assuming a constant number of runs  of zeros, label the runs by integers modulo $m$  and consider block codes over the ring of integers modulo $m$---the smallest possible $m$ depends on $d$ and $k$.

Our approach differs from the one in \cite{LH} in two ways. 
First, we relax the unit length runlength of the ones in \cite{LH} (which was motivated by magnetic recording applications). Second, we consider
lattices rather than codes over  the integers modulo $m$ to allow a wider choice of parameters. Indeed our deletion codes are
obtained as sets of vectors in a lattice with a given Manhattan norm. By varying this norm, a single lattice, possibly obtained from a single Lee code by Construction A,  can produce an infinity of deletion codes.
We extend some results
of \cite{B,So} on generalized theta series, called there $\nu-$series, to effectively enumerate these special sets of vectors in the lattice.
In particular, if the lattice is obtained via Construction A from a code, the generalized $\nu-$series allows to enumerate these sets
from the weight enumerators of the code.

The paper is organized as follows. In Section~\ref{one}, we formalize the problem. 
In Section~\ref{two}, we determine the sizes of codes derived from Construction $A$ lattices.  In Section~\ref{three} we provide a codebook generation algorithm and a corresponding decoding algorithm for a specific class of lattices which includes the $E_8$ lattice. 
In Section~\ref{four}, using tools developed in Section~\ref{two} we derive the analogue of the Gilbert and Hamming bounds for the Manhattan metric space.  In Section~\ref{five} we derive the asymptotic versions of these bounds.
In Section~\ref{six}, we provide a few concluding remarks and point to some open problems.
\section{Background and Statement of the Problem}\label{one}
Consider a binary sequence of length $N$ that starts with a zero and that contains an even number $n$ of runs---hence $n/2$ runs of zeros and $n/2$ runs of ones. For instance,  the sequence $0011100011$ corresponds to $N=10$ and $n=4$. Throughout the paper we make the following hypothesis: 
\begin{WH} In any given code $n$ is the same across codewords and they all start with a zero. Moreover, the runlengths in each  
codeword are supposed to be lower bounded by some constant $r\geq 1$ where $r-1$ corresponds to the maximum number of deletions that can occur over a length $N$ codeword. This condition is imposed so that the number of runs before and after transmission remains the same.
\end{WH}

With a given length $N$ binary sequence we associate its corresponding runlength sequence 
$$(x_1,y_1,\dots,x_i,y_i,\dots,x_{n/2},y_{n/2})$$
where $x_i$ and $y_i$ denote the $i$\/th runlength of zeros and ones, respectively. 
For instance, sequence $0011100011$ corresponds to $(2,3,3,2).$ The integer sequence so constructed satisfies the constraint
$$N=\sum_{i=1}^{n/2}(x_i+y_i).$$

Denote by $\phi$ the above correspondence from $\F_2^N$ to $\Z^{n}.$
The {\bf Levenshtein distance} between two binary vectors is the least number of deletions to go from one to the other \cite{Lev}.
The {\bf Manhattan distance} between two vectors ${\bf w,z} \in \Z^{n}$ is defined as

$$|{\bf w}-{\bf z}|\defeq \sum_{i=1}^n|w_i-z_i|.$$
The following observation is trivial but crucial.

{\prop Under the above working hypothesis, the map $\phi$ is an isometry between $\F_2^N$ with the Levenshtein distance and $\Z^{n}$ with
the Manhattan distance.}
\begin{proof}
Let $${\bf z}=(x_1,y_1,\cdots,x_n,y_n)$$ denote a sequence of runs. Let $j$ be an integer $\le r-1.$ Any deletion of $j$ zeros (resp. ones) into run number $i$
will result into a change of $x_i$ (resp. $y_i$) into $x_i\pm j$ (resp. $y_i\pm j$) yielding a sequence ${\bf z'}$ at Manhattan distance $j$ away from ${\bf z}.$\end{proof}
The problem we consider is to characterize $A(n,d,N,r)$, the largest number of length $n$ vectors of nonnegative integers at Manhattan distance at least $d$ apart 
and with 
coordinates summing up to $N.$ 
Any set of length $n$ vectors with integral entries $\ge r$, at Manhattan distance at least $d$ apart, and coordinates summing up to $N,$ we refer to as an $(n,d,N,r)-$set.
%%%%%%%%%%%%%%%%%%%%%%%%%%%%%%%%%%%%%%%%%%%%%%%%%%%%
\section{ Enumeration for construction A lattices}\label{two}
A  {\bf code} $C\subseteq \Z_m^n$ is defined as a $\Z_m-$submodule of $\Z_m^n.$
The {\bf complete weight enumerator} (cwe) of $C$ is defined as
the polynomial (see \cite[Chap. 5.6]{SMW})
$$cwe_C(x_1,x_2,\hdots, x_m)=\sum_{c \in C} \prod_{i=0}^{m-1}x_i^{n_i(C)},$$
where $n_i(c)$ is the number of entries equal to $i$ in the vector $c.$
For $m=2$, we let $$W_C(x,y)\defeq cwe_C(x,y)$$ be the classical {\bf weight enumerator} of a binary code. 

A  {\bf lattice} of $\R^n$ is defined as a discrete additive subgroup of $\R^n.$ 
A lattice $L$ is said to be obtained by {\bf Construction~A}
from a code $C$ of $\Z_m^n$ if $C$ is the image of $L$ by reduction modulo $m$ componentwise \cite[Chap. 7.2]{CS}. 
Such a lattice is denoted by $L=A(C).$ An important parameter of a lattice is its minimum distance (norm) which is given by the following proposition. 
Recall that the {\bf Lee weight}  of a symbol $x\in \Z_m=\{0,1,\cdots,m-1\}$ is defined as $$\min(x, m-x).$$ The weight of a vector is the sum of the weights of its components, 
and the Lee distance of two vectors is the Lee weight of their difference vector. The {\bf Lee distance} of a linear code
$C \subseteq \Z_m^n$ is the minimum weight of its nonzero elements.

{\prop [\cite{RS}]\label{prop:RS} Let $L=A(C)$ for some $C\subseteq \Z_m^n$. Then the minimum distance of $L$ is given by $$d=\min(d',m)$$}
where $d'$ is the minimum Lee distance of $C$.

For an integer $r\ge 0$ define
$$\nu_L(r;q)\defeq \sum_{\substack{{\bf x} \in L:\\ \min_i x_i \ge r }} q^{|{\bf x}|} $$ as
the shifted $\nu-$series in the indeterminate $q$ of the lattice~$L$.

This definition extends trivially to any discrete subset $L$ of $\R^n.$
The motivation for this generating function, whose case $r=0$ is the $\nu-$series of \cite{B,S}, stems
from Proposition~\ref{prop:maxsize} below which gives a lower bound on $A(n,d,N,r)$. 
\begin{nota} We use the Waterloo notation for coefficients of generating series (see \cite{GJ}).
 Given $q-$series  $f=\sum_i f_i q^ i$ we
denote by $[q^i]f(q)$ the coefficient $f_i.$
\end{nota}
{\prop \label{prop:maxsize} If $L$ is a lattice of $\R^n$ with minimum Manhattan distance $d$ then 
the set of vectors of $L$ with coordinate entries bounded below by $r$
and Manhattan norm $N$
forms an $(n,d,N,r)-$set of size $[q^N]\nu_L(r;q) \le A(n,d,N,r).$
}

The proof of Proposition~\ref{prop:maxsize} immediately follows from the definition of $[q^N]\nu_L(r;q)$ and $A(n,d,N,r)$.
%We denote by $W_C(x,y)$ the weight enumerator of $C$ for $m=2,$ and by $cwe_C(a,b,c)$ the complete weight enumerator of $C$ for $m=4.$

We now show how to compute (shifted) $\nu-$series of lattices from (complete) weight enumerators of codes.

{\thm If $L=A(C)$ and $m=2$ then  $$\nu_L(r;q)=W_C(\frac{q^a}{1-q^2},\frac{q^b}{1-q^2}),$$
where $a$ (resp. $b$) is the first even (resp. odd) integer $\ge r.$
If $L=A(C)$ and $m=4,$ then  $$\nu_L(r;q)=cwe_C(\frac{q^a}{1-q^4},\frac{q^b}{1-q^4},\frac{q^c}{1-q^4},\frac{q^d}{1-q^4}),$$
where $a,b,c,d$ are the first integers $\ge r,$ congruent to $0,1,2,3$ modulo $4$ respectively.
  }

\begin{proof}
Use the same argument as in \cite{B,So} and write $A(C)$ as a disjoint union of cosets of $m\Z^n$
 $$\nu_L(r;q)=W_C(\nu_{2\Z}(r;q),\nu_{2\Z+1}(r;q))$$
 for $m=2,$
and 
$$\nu_L(r;q)=cwe_C(\nu_{4\Z}(r;q),\nu_{4\Z+1}(r;q),\nu_{4\Z+2}(r;q),\nu_{4\Z+3}(r;q))$$
for $m=4$, respectively. The result follows by observing that
$$\nu_{4\Z}(r;q)=\frac{q^a}{1-q^4}$$
and by summing the appropriate geometric series of reason $q^2$ or $q^4.$
\end{proof}

In Column 2 of Tables \ref{table:E82}, \ref{table:E84}, and \ref{table:BW16L24},  we list for some values of $N$ and $r$ 
the lower bound $[q^N]\nu_L(r;q)$ to $A(n,d,N,r)$ for the well-known lattices $E_8$, $BW_{16}$, and $\Lambda_{24}$. These lattices are constructed from the extended Hamming code $H_8$ modulo $2$ or the Klemm code $K_8$ modulo $4$ for $E_8$, the code $RM(1,4)+2RM(2,4)$ for $BW_{16}$, 
and the lifted Golay code ${\mathcal QR}_{24}$ for $\Lambda_{24}$. Here $K_s=R_s+2P_s$ where $R_s$ denotes the length$-s$ repetition code, where$P_s=R_s^{\perp}$ denotes its dual code, and where $RM(k,m)$ denotes the order-$k$ Reed-Muller code of length $2^m.$ 

Some cwe's for these codes can be found in \cite{BSBM,BSC} while others were computed using Magma \cite{mag}. The cwe of $K_n$ is easily seen to be
$$ \frac{1}{2}[(x_0+x_2)^n+(x_0-x_2)^n+(x_1+x_3)^n+(x_1-x_3)^n].$$
These numerical results show, for instance, that for $r=2$ and $N=64$, among the three lattices $ E_8$, $BW_{16}$ and $\Lambda_{24}$, $BW_{16}$ achieves the best lower bound while $\Lambda_{24}$ achieves the best bound for $r=1$ and $N=64$. 

We now add an extra ingredient to the above construction which improves the lower bound on $A(n,d,N,r)$ for $N$ large enough. Let $L$ be a Construction A lattice in $\Z^{n-1}$ with $L^1-$distance $d$. From this lattice in $\Z^{n-1}$ we construct a new set of points in $\Z^{n}$ as
$$\widehat{L}\defeq \{(x_1,x_2,\hdots,x_{n-1},N-\sum\limits_{i=1}^{n-1}x_i)|(x_1,\hdots,x_{n-1})\in L\}.$$
Note that the map $$(x_1,x_2,\hdots,x_{n-1})\mapsto (x_1,x_2,\hdots,x_{n-1},N-\sum\limits_{i=1}^{n-1}x_i)$$
is the Manhattan analogue map of the Yaglom map (see, {\it{e.g.}}, \cite[Chap. 9, Theorem~6]{CS})
$$(x_1,x_2,\hdots,x_{n-1})\mapsto (x_1,x_2,\hdots,x_{n-1},({N^2-\sum\limits_{i=1}^{n-1}x_i^2})^{1/2})$$
from ${\mathbb{R}}^{n-1}$ to ${\mathbb{R}}^n$.

Column $3$ of Tables \ref{table:E82} and \ref{table:E84} gives the lower bound $[q^N]\nu_{\hat{L}}(r;q)$ for the secondly proposed code construction. As we can observe, for $N$ large enough ({\it{e.g.}}, $N\geq 28$ for $E_8$), this second construction improves the first.
\begin{table}[ht]
\centering \caption{\label{table:E82}Size $[q^N]\nu_L(r;q)$ of $(n,d,N,r)-$ set with $L=A(H_8), d\geq 2$ and  $r=1,2$}
%\setlength{\columnsep}{0.005cm}
%\begin{multicols}{2}
%\scriptsize
%\tiny
\begin{tabular}{|ccc|}
\hline
$N$&$[q^N]\nu_{E_8}(1;q)$&$[q^N]\nu_{\widehat E_8}(1;q)$\\

8&1&0\\10&8&1 \\ 12&50&9 \\ 14&232&59 \\ 16&835&291 \\ 18&2480&1126 \\ 20&6372&3606 \\ 22&14640&9978 \\ 
24&30789&24618 \\ 26&60280&55407 \\ 28&111254&115687 \\ 30&195416&226941 \\ 32&329095&422357 \\ 
34&534496&751452 \\ 36&841160&1285948\\
\hline
$N$&$[q^N]\nu_{E_8}(2;q)$&$[q^N]\nu_{\widehat E_8}(2;q)$\\
16&1&0\\18&8&1 \\20&50&9 \\22&232&59 \\24&835&291 \\26&2480&1126 \\28&6372&3606\\
30&14640&9978 \\32&30789&24618 \\34&60280&55407 \\36&111254&115687 \\
38&195416&226941 \\40&329095&422357 \\42&534496&751452 \\44&841160&1285948 \\
\hline
\end{tabular}
\end{table}

\begin{table}[ht]
\centering \caption{\label{table:E84}Size $[q^N]\nu_L(r;q)$ of $(n,d,N,r)-$ set with $L=A(K_8), d\geq 4$ and  $r=1,2$}
\begin{tabular}{|ccc|}
\hline
$N$&$[q^N]\nu_{E_8}(1;q)$&$[q^N]\nu_{\widehat E_8}(1;q)$ \\
 		8&			1&			0\\         
 		12&        36&        1    \\    
        16&        331&        37    \\

        20&        1752&        368    \\
    
        24&        6765&        2120    \\
    
        28&        21164&        8885    \\
    
        32&        56823&        30049    \\
    
        36&        135728&        86872    \\
    
        40&        295545&        222600    \\
    
        44&        596980&        518145    \\
    
        48&        1133187&        1115125    \\    
        52&        2041480&        2248312    \\
    
        56&        3517605&        4289792    \\
    
        60&        5832828&        7807397    \\    
        64&        9354095&        13640225    \\
\hline
$N$&$[q^N]\nu_{E_8}(2;q)$&$[q^N]\nu_{\widehat E_8}(1;q)$\\

		16&			1&			0\\    
		20&        36&        1    \\
    
        24&        331&        37    \\
    
        28&        1752&        368    \\
    
        32&        6765&        2120    \\
    
        36&        21164&        8885    \\
    
        40&        56823&        30049    \\
    
        44&        135728&        86872    \\
    
        48&        295545&        222600    \\
    
        52&        596980&        518145    \\
    
        56&        1133187&        1115125    \\
    
        60&        2041480&        2248312    \\
    
        64&        3517605&        4289792    \\
    
        68&        5832828&        7807397    \\
    
        72&        9354095&        13640225    \\

\hline
\end{tabular}

%\end{multicols}
\end{table}

\begin{table}[ht]
\centering \caption{\label{table:BW16L24}Size $[q^N]\nu_L(r;q)$ of $(n,d,N,r)-$ set with $L=BW_{16},\Lambda_{24}, d\geq 4$ and $r=1,2$ }
\setlength{\columnsep}{0.005cm}
\begin{multicols}{2}
%\scriptsize
%\tiny
\begin{tabular}{|cc|}
\hline
$N$&$[q^N]\nu_{BW_{16}}(1;q)$\\
    
        16&1\\
        20&16\\
        24&306\\
        28&3984\\
        32&39235\\
        36&310176\\
        40&2016996\\
        44&11005344\\
        48&51463749\\
        52&210557360\\
        56&767796630\\
        60&2535136560\\
        64&7680579975\\
        68&21588192576\\
        72&56814408136\\
        %76&141077361984\\
        %80&332674600329\\
        %84&749069927760\\
        %88&1618051018746\\
        %92&3366425276624\\
\hline

$N$&$[q^N]\nu_{BW_{16}}(2;q)$\\

    32&1\\
    36&16\\
    40&306\\
    44&3984\\
    48&39235\\
    52&310176\\
    56&2016996\\
    60&11005344\\
    64&51463749\\
    68&210557360\\
    72&767796630\\
    76&2535136560\\
    80&7680579975\\
    84&21588192576\\
    88&56814408136\\
\hline

\end{tabular}

\begin{tabular}{|cc|}
\hline
$N$&$[q^N]\nu_{\Lambda_{24}}(1;q)$\\
    
    24&1\\
    28&24\\
    32&300\\
    36&2600\\
    40&23415\\
    44&299760\\
    48&4144211\\
    52&48058824\\
    56&448956690\\
    60&3450990152\\
    64&22448210613\\
    68&126639274800\\
    72&632120648146\\
    76&2837407970784\\
    80&11605964888130\\
    \hline

$N$&$[q^N]\nu_{\Lambda_{24}}(2;q)$\\

    48&1\\
    52&24\\
    56&300\\
    60&2600\\
    64&23415\\
    68&299760\\
    72&4144211\\
    76&48058824\\
    80&448956690\\
    84&3450990152\\
    88&22448210613\\
    92&126639274800\\
    96&632120648146\\
    100&2837407970784\\
    104&11605964888130\\
\hline

\end{tabular}

\end{multicols}
\end{table}

In this section we derived lower bounds on $A(n,d,N,r)$ in a non-constructive fashion from the properties of $L$ and $\hat{L}$ using generating functions (Proposition~\ref{prop:maxsize}). In the next section we provide an explicit code construction for a specific family of lattices along with an effective decoding algorithm.

%%%%%%%%%%%%%%%%%%%%%%%%%%%%%%%%%%%%%%%%%%%%%%%%%%%%
\section{Code construction and decoding algorithm}\label{three}
In this section, we describe two algorithms with respect to the lattice $A(K_n)$:
\begin{itemize}
 \item a search algorithm that generates explicitly an $(n,N,d,r)$ set carved from the lattice;
\item a corresponding decoding algorithm.
\end{itemize}

Define code $$C(n,d,N,r)\defeq \{{\bf c} \in A(K_n):\min\limits_i c_i \ge r, \: \sum\limits_{i=1}^nc_i=N\}$$ and note that
the minimum distance of $C(n,d,N,r)$ is at least $4$, the minimum distance inherited from $ A(K_n)$.
The generator matrix $G $ for the lattice $A(K_n)$ is
$$G=\left(
\begin{array}{cccccc}
1 & 1 & 1 & \cdots & 1	& 1\\
0 & 2 & 0 & \cdots & 0	& 2\\
0 & 0 & 2 & \cdots & 0	& 2\\
\vdots & \vdots & \vdots & \ddots & \vdots	& \vdots\\
0 & 0 & 0 & \cdots & 2	& 2\\
0 & 0 & 0 & \cdots & 0	& 4\\
\end{array}
\right)$$
hence any codeword ${\bf c}$ in $C(n,d,N,r)$ can be expressed as 
$${\bf c}=(x_1,x_1+2x_2,\hdots,x_1+2x_{n-1},x_1+2\sum\limits_{i=2}^{n-1}x_i+4x_n)$$ with $$l_i\leq x_i\leq u_i$$ and where $l_i$ and $u_i$
are determined as follows. 

Define $$S_i\defeq x_1+\sum\limits_{j=2}^{i}(x_1+2x_j)$$ and $$T\defeq x_1+2\sum\limits_{j=2}^{n-1}x_j.$$ Then 
\begin{itemize}
	\item for $i=1,$
\begin{equation*}
\begin{array}{l}
l_1=r\\
u_1=N-(n-1)r,
\end{array}
\end{equation*}

	\item for $2\leq i\leq n-1$, 

\begin{equation*}
\begin{array}{l}
l_i=\left\lceil\frac{1}{2}\left(r-x_1\right)\right\rceil\\
u_i=\left\lfloor\frac{1}{2}\left(N-(n-i)r-S_{i-1}-x_1\right)\right\rfloor,
\end{array}
\end{equation*}

\item for $i=n,$
\begin{equation*}
\begin{array}{l}
l_n=\left\lceil {\frac{1}{4}\left(r-T\right)}\right\rceil\\
u_n=\left\lfloor \frac{1}{4}\left(N-(n-1)r-T\right)\right\rfloor.\\
\end{array}
\end{equation*}	
\end{itemize}
Searching the codewords can be done by a tree search through all nodes from level $1$ (corresponding to $x_1$) to level $n$ (corresponding to $x_n$). With the above constraints, we are able to efficiently generate all codewords in $C(n,d,N,r)$. Numerical results are given in Table \ref{table:RLL}.

Table \ref{compare:search_algo_N12} gives for $n=8$, $N=12$, $r=1$ and the quaternary lattice $E_8=A(K_8)$ the number of visited nodes at level $i$ and its naive upper bound which is roughly $(N-7)(\frac{N-6}{2})^{i-1}$, for different $i$'s. Table \ref{compare:search_algo} gives the number of visited nodes at level $i=6$ for different values of $N$ (we keep $n=8$ and $r=1$).

\begin{table}[ht]
\centering \caption{\label{table:RLL}Codewords in $E_8$ with RLL representation for $r=1,N=12$}
\setlength{\columnsep}{0.005cm}
\begin{multicols}{2}
%\scriptsize
%\tiny
\begin{tabular}{|c|}
\hline
    $( 5, 1, 1, 1, 1, 1, 1, 1 )$\\
    $( 3, 1, 1, 1, 1, 1, 1, 3 )$\\
    $( 3, 1, 1, 1, 1, 1, 3, 1 )$\\
    $( 3, 1, 1, 1, 1, 3, 1, 1 )$\\
    $( 3, 1, 1, 1, 3, 1, 1, 1 )$\\
    $( 3, 1, 1, 3, 1, 1, 1, 1 )$\\
    $( 3, 1, 3, 1, 1, 1, 1, 1 )$\\
    $( 3, 3, 1, 1, 1, 1, 1, 1 )$\\
    $( 1, 1, 1, 1, 1, 1, 1, 5 )$\\
    $( 1, 1, 1, 1, 1, 1, 3, 3 )$\\
    $( 1, 1, 1, 1, 1, 1, 5, 1 )$\\
    $( 1, 1, 1, 1, 1, 3, 1, 3 )$\\
    $( 1, 1, 1, 1, 1, 3, 3, 1 )$\\
    $( 1, 1, 1, 1, 1, 5, 1, 1 )$\\
    $( 1, 1, 1, 1, 3, 1, 1, 3 )$\\
    $( 1, 1, 1, 1, 3, 1, 3, 1 )$\\
    $( 1, 1, 1, 1, 3, 3, 1, 1 )$\\
    $( 1, 1, 1, 1, 5, 1, 1, 1 )$\\
    \hline
 \end{tabular}
 
 \begin{tabular}{|c|}
 \hline
    $( 1, 1, 1, 3, 1, 1, 1, 3 )$\\
    $( 1, 1, 1, 3, 1, 1, 3, 1 )$\\
    $( 1, 1, 1, 3, 1, 3, 1, 1 )$\\
    $( 1, 1, 1, 3, 3, 1, 1, 1 )$\\
    $( 1, 1, 1, 5, 1, 1, 1, 1 )$\\
    $( 1, 1, 3, 1, 1, 1, 1, 3 )$\\
    $( 1, 1, 3, 1, 1, 1, 3, 1 )$\\
    $( 1, 1, 3, 1, 1, 3, 1, 1 )$\\
    $( 1, 1, 3, 1, 3, 1, 1, 1 )$\\
    $( 1, 1, 3, 3, 1, 1, 1, 1 )$\\
    $( 1, 1, 5, 1, 1, 1, 1, 1 )$\\
    $( 1, 3, 1, 1, 1, 1, 1, 3 )$\\
    $( 1, 3, 1, 1, 1, 1, 3, 1 )$\\
    $( 1, 3, 1, 1, 1, 3, 1, 1 )$\\
    $( 1, 3, 1, 1, 3, 1, 1, 1 )$\\
    $( 1, 3, 1, 3, 1, 1, 1, 1 )$\\
    $( 1, 3, 3, 1, 1, 1, 1, 1 )$\\
    $( 1, 5, 1, 1, 1, 1, 1, 1 )$\\
    \hline
\end{tabular}
\end{multicols}
\end{table}

\begin{table}[ht]
\centering \caption{\label{compare:search_algo_N12} Number of visited nodes and its upper bound of searching codewords from $E_8=A(K_8)$ with  $r=1,N=12$}
%\setlength{\columnsep}{0.005cm}
%\begin{multicols}{2}
%\scriptsize
%\tiny
\begin{tabular}{|ccc|}
\hline
Level &$\#$nodes&Upper bound \\

		2&			9&15\\         
 		3&        11&45    \\    
        4&        16&135    \\

        5&        21&405 \\
    
        6&        28&1215 \\
    
        7&        36&3645    \\
    
\hline
\end{tabular}
\end{table}

\begin{table}[ht]
\centering \caption{\label{compare:search_algo} Number of visited nodes and its upper bound of searching codewords from $E_8=A(K_8)$ for  $r=1$}
%\setlength{\columnsep}{0.005cm}
%\begin{multicols}{2}
%\scriptsize
%\tiny
\begin{tabular}{|cccc|}
\hline
%$N$&$[q^N]\nu_{E_8}(1;q)$&Nb. of visited nodes&Upper bound \\
%
%		8&			1&			1&1\\         
% 		12&        36&        36&3645    \\    
%        16&        331&        338&140625    \\
%
%        20&        1752&        1836&1529437 \\
%    
%        24&        6765&        7277&9034497 \\
%    
%        28&        21164&        22880&37202781    \\
%    
%        32&        56823&        61828&120670225    \\
%    
%        36&        135728&        148104&330328125    \\
%    
%        40&        295545&        322677&796539777    \\
%    
%        44&        596980&        651244&1740697597    \\
%    
%        48&        1133187&        1234134 &3516410961   \\    
%        52&        2041480&        2218580&6661615005    \\
%    
%        56&        3517605&        3813615 &11962890625   \\
%    
%        60&        5832828&        6307848&20533285917    \\    
%        64&        9354095&10090376 &33904929297           \\

$N$&$\#$nodes(level 7)&$\#$nodes(level 6)&Upper bound(level 6) \\

		8&			1&	1&		1\\         
 		12&        36&    28&    1215    \\    
        16&        331&   217&     28125    \\

        20&        1752&   1008&     218491 \\
    
        24&        6765&    3465&    1003833 \\
    
        28&        21164&    9724&    3382071    \\
    
        32&        56823&     23569&   9282325    \\
    
        36&        135728&    51136&    22021875    \\
    
        40&        295545&     101745&   46855281    \\
    
        44&        596980&     188860&   91615663    \\
    
        48&        1133187&    331177&    167448141   \\    
        52&        2041480&    553840 &   289635435    \\
    
        56&        3517605&    889785&    478515625   \\
    
        60&        5832828&    1381212&    760492071    \\    
        64&        9354095&		2081185&	1169135493           \\
\hline
\end{tabular}
\end{table}

%Before going to the decoding problem, let us recall some result of Campello et al. \cite{CJC}. In \cite{CJC}, they gave a decoding algorithm for a Construction A $q-$ary lattice whose $q-$ary linear code associated has its generator matrix written in the systematic form and the efficiency of their algorithm strongly depends on the dimension of the linear code. Unfortunately, in our case, the associated Klemm code $K_n$ of length $n$ makes their decoding algorithm less efficient due to the process in the exhaustive search.

We now turn to decoding. Recall that in \cite{CJC} the decoding of a Construction A $q-$ary lattice for the $L^1-$norm is reduced to that of a $q-$ary linear code for the Lee metric.

We now describe our decoding algorithm for the $C(n,N,d,r)$ code (carved from $A(K_n)$) using the runlength limited (RLL) sequence of its codewords. Recall that, because of our working hypothesis, the channel preserves the number of runs. 

From the definition of $A(K_n)$ we have $$A(K_n)=2D_n\cup ({\bf 1}+2D_n),$$
where $$D_n\defeq \{{\bf x}\in \Z^n|\sum\limits_{i=1}^nx_i\equiv 0\mod 2\}.$$
It is clear that $D_n$ contains $$A_{n-1}=\{{\bf x}\in \Z^n|\sum\limits_{i=1}^nx_i=0\}$$ as a sublattice. 

Following \cite{CS82}, we reduce the decoding in $2D_n$ to the decoding in $2A_{n-1}$ by noting that
$$2D_n=k+2A_{n-1}$$
with $k=(N,0,\ldots,0)$.

The following lemma allows us to find a closest codeword in $A_{n-1}$ to a received vector in ${\mathbb Z}^n.$
{\lem \label{lem:nearest point} Any vector of coordinates summing up to $s$ in $\Z_+^n$ is at $L^1-$distance at least $|s|$ from any vector in $A_{n-1}$.}
\begin{proof}
Let ${\bf x}=(x_1,x_2,\hdots,x_n)\in \Z_+^n$ with $\sum\limits_{i=1}^n x_i=s$ and ${\bf y}\in A_{n-1}.$ 
Then 
$${|\bf x-y|}=\sum\limits_{i=1}^n|x_i-y_i|\ge |\sum\limits_{i=1}^n(x_i-y_i)|= s$$ 
since $\sum_{i=1}^n y_i=0$.
\end{proof}

{\prop Let 
\begin{align}
\phi^{(i)}:\Z^n&\rightarrow A_{n-1}\\ {\bf x}&\mapsto (\phi_1^{(i)},\phi_2^{(i)},\hdots,\phi_n^{(i)}),
\end{align}
where \begin{equation*} 
\phi_j^{(i)}=
\begin{cases}
x_j-(x_1+\cdots+x_n) & \text {if } j=i\\
x_j & \text {if } j\neq i.\\
\end{cases}
\end{equation*}
Then for any ${\bf x}\in \Z_+^n$, $\phi^{(i)}({\bf x})$ is a closest point of $A_{n-1}$ to~${\bf x}$.
}
\begin{proof} 
The proof follows from Lemma \ref{lem:nearest point} with $s=|{\bf x}|$.
\end{proof}
In case of a single deletion error (recall that the minimum distance of $A(K_n)$ is $4$), there exists a unique $i\in\{1,2,\hdots,n\}$ such that $2A_{n-1}$ contains $\phi^{(i)}({\bf x})$. That $i$ is where the error occurs.

\noindent \rule[-.1cm]{\linewidth}{0.2mm}
{\bf Algorithm}\\
\noindent \rule[.3cm]{\linewidth}{0.2mm}

\vspace{-.3cm}
\noindent\textbf{Input:} A received vector ${\bf x}$ of length $n$\\
\textbf{Output:} A nearest codeword $\hat{\bf x} $ to $\bf {x}$
%\vspace{.2cm}
\begin{enumerate}
\item $N\leftarrow$ length of the binary code corresponding
\item ${\bf a}\leftarrow$ a coset representative of $A_{n-1}$ in $D_{n}$
\item if $\sum\limits_{i=1}^n{\bf x}[i]==N-1$ then
\item $\hat{\bf x}\leftarrow \bf {x}$
\item Find (the unique) coordinate $\hat{\bf x}[j]$ whose parity is different from the others
\item $\hat{{\bf x}}[j]\leftarrow \hat{{\bf x}}[j]+1$
\item else
\item $\hat{\bf X}\leftarrow \bf {x-a}$
\item $s\leftarrow\sum\limits_{i=1}^n{{\hat X}}[i]$
\item for $i\leftarrow 1$ to $n$ do
\item $\hat {\bf x}\leftarrow \hat {\bf X}$
\item ${\hat {\bf x}}[i]\leftarrow {\hat {\bf x}}[i]-s$
\item if all coordinates of ${\hat {\bf x}}$ are even then 
\item break
\item end if
\item end for
\item $\hat{{\bf x}}\leftarrow \hat{{\bf x}}+{\bf a}$
%\item if $\sum\limits_{i=1}^n{\hat {\bf x}}[i]<N$ then
%\item ${\hat {\bf x}}\leftarrow {\hat {\bf x}}+{\bf 1}$
%\item end if
\item end if
\item return $\hat{\bf x}$
\end{enumerate}
\noindent \rule[.3cm]{\linewidth}{0.2mm}\\
The complexity of our algorithm can be calculated as follows:
\begin{itemize}
\item line $3$ requires $n-1$ additions
\item line $8$ requires $n$ additions
\item line $9$ requires $n-1$ additions
\item lines $10$ to $16$ require one addition (plus one parity test) for $n$ times 
\item line $17$ requires $n$ additions
\end{itemize}
Thus the decoding algorithm requires $5n-2$ additions over $\mathbb{Z}$ plus $n$ parity tests.

For instance, take $n=8,N=12,r=1$ and consider ${\bf x}=(3,2,1,1,1,1,1,1)$ as a received word. 
The code $C(8,12,1)$ has $36$ codewords and has minimum distance $4$. By taking as coset representative of $A_{n-1}$ in $D_n$ $${\bf a}=(1,1,1,1,1,1,1,5),$$ the nearest codewords in $A_{n-1}$ to ${\bf x-a}$ are 
$$\phi^{(1)}({\bf x-a})=( 2, 1, 0, 0, 0, 0, 0, -4 ),$$
$$\phi^{(2)}({\bf x-a})=( 2, 2, 0, 0, 0, 0, 0, -4 ),$$
$$\phi^{(3)}({\bf x-a})=( 2, 1, 1, 0, 0, 0, 0, -4 ),$$
$$\phi^{(4)}({\bf x-a})=( 2, 1, 0, 1, 0, 0, 0, -4 ),$$
$$\phi^{(5)}({\bf x-a})=( 2, 1, 0, 0, 1, 0, 0, -4 ),$$
$$\phi^{(6)}({\bf x-a})=( 2, 1, 0, 0, 0, 1, 0, -4 ),$$
$$\phi^{(7)}({\bf x-a})=( 2, 1, 0, 0, 0, 0, 1, -4 ),$$
$$\phi^{(8)}({\bf x-a})=( 2, 1, 0, 0, 0, 0, 0, -3 ).$$
Since $\phi^{(2)}({\bf x-a})$ is the only codeword in $2A_{n-1}$, we decode ${\bf x}=(3,2,1,1,1,1,1,1)$ since $$\phi^{(2)}({\bf x-a})+{\bf a}=(3,3,1,1,1,1,1,1).$$

%%%%%%%%%%%%%%%%%%%%%%%%%%%%%%%%%%%%%%%%%%%%%%%%%%%%
\section{Bounds on $A(n,d,N,r)$}\label{four}
 First we recall a well-known identity of  formal power series.
{\lem \label{wk} For any integer $n\ge 1$, we have $$\frac{1}{(1-q)^n}=\sum_{i=0}^\infty {{i+n-1}\choose {n-1}}q^i.$$}
\begin{proof}
Differentiate the geometric series  $$ \frac{1}{(1-q)}=\sum_{i=0}^\infty q^ i$$ with respect to $q$ and use induction on $n.$
\end{proof}
Using generating functions, we compute the volume $V(n,e)$ of the Manhattan ball of radius $e$ in $\Z^n.$

{\lem \label{2} For any integers $n\ge e\ge 1$, we have $$V(n,e)=[q^e]\frac{(1+q)^n}{(1-q)^{n+1}}=\sum_{i=0}^{\min(n,e)}
2^i {{n}\choose {i}}{{e}\choose {i}}.$$}
\begin{proof}
\begin{align*}
V(n,e)&=\sum_{i=0}^e[q^i]\nu_{\Z^n}(-\infty,q)\\
&=\sum_{i=0}^e[q^i](\frac{1+q}{1-q})^n\\
&=[q^e]\frac{(1+q)^n}{(1-q)^{n+1}}.
\end{align*}
%The result follows by Lemma \ref{wk}.
The second expression in the Lemma is from \cite{GW}. It can be rederived from the above generating series by expanding
$$(1+\frac{2q}{1-q})^{n+1} =\sum_{i=0}^n {n \choose i}2^i \frac{q^i}{(1-q)^{i+1}}$$
through Lemma \ref{wk}.
\end{proof}
By the same techniques, we can compute the volume of the ambient space $A(n,1,N,r).$

{\lem \label{3}  For any integer $N> nr$ and $r>e\ge 1$, we have $$A(n,1,N,r)={{N-nr+n-1}\choose {n-1}}.$$}

\begin{proof}
\begin{tabbing}
$A(n,1,N,r)$ \=$=[q^N]\nu_{\Z^n}(r,q)=[q^N](q^r\frac{1}{1-q})^n$\\
\>$=[q^{N-nr}]\frac{1}{(1-q)^n}.$
\end{tabbing}
%$$A(n,1,N,r)=[q^N]\nu_{\Z^n}(r,q)=[q^N](q^r\frac{1}{1-q})^n=[q^{N-nr}]\frac{1}{(1-q)^n}.$$
The result follows from Lemma \ref{wk}.
\end{proof}

We are now in a position to formulate the analogues of the Gilbert and Hamming bound in the present context.

{\thm \label{fonda} For any integers $N> nr, n\ge d$, and $r>e=\lfloor (d-1)/2\rfloor \ge 1$,  we have
$$     \frac{ {{N-nr+n-1}\choose {n-1}}  }{V(n,d-1)}        \le   A(n,d,N,r)\le \frac{ {{N-nr+n-1}\choose {n-1}}  }{V(n,e)}               .$$
 }

\begin{proof}
Combine Lemma \ref{2} and Lemma \ref{3} with the standard arguments.
\end{proof}
%The numerical results of some well-known lattices for these bounds 
The lower and upper bounds on $A(n,d,N,r)$ in Theorem~\ref{fonda} 
are given  in Table \ref{table3} and Table \ref{table4} for lattices $E_8$ and $BW_{16}$. In these tables we defined 
$$I(n,d,N,r)\defeq \left\lceil\frac{{N-nr+n-1 \choose n-1}}{V(n,d-1)}\right\rceil$$ and 
$$S(n,e,N,r)\defeq \left\lceil\frac{{N-nr+n-1 \choose n-1}}{V(n,d-1)}\right\rceil .$$ The numerical results show
that  $[q^N]\nu_L(r;q)$ (a lower bound to $A(n,d,N,r)$ by Proposition \ref{prop:maxsize}),  lies between $I(n,d,N,r)$ and $S(n,e,N,r)$ for many parameter values. Exceptions are, for instance, for $BW_{16}$ with $r=2$, and $N=48,\hdots,96$. Whether these code constructions yield sizes between $I(n,d,N,r)$ and $S(n,e,N,r)$ for large $N$ is an open issue.

Since all codewords have constant Manhattan distance, it is natural to consider the Johnson bound in the Lee metric:
{\thm \label{johnson} If $d>N(1-1/2n),$ then we have $$A(n,d,N,r)\le \frac{d}{d-N(1-1/2n)}.$$}
\begin{proof}
Reduce all vectors modulo $Q=2N.$
Use Lemma 13.62 of \cite{Be} with $\overline{D}=Q/4=N/2,$ and $x=1/n.$
\end{proof}

\begin{table}[ht]
\centering \caption{\label{table3} Bounds on $A(n,d,N,r)$ with $L=E_8$ and $r=2,3,4$}
%\setlength{\columnsep}{0.05cm}
%\begin{multicols}{2}
\scriptsize
{
%\scriptsize{
\begin{tabular}{|cccc|}
\hline
$N$&$I(8,4,N,2)$&$[q^N]\nu_{E_{8}}(2;q)$&$S(8,1,N,2)$\\
%\hline
          
    %16&1&1&0\\
    %20&1&36&19\\
    24&8&331&378\\
    28&61&1752&2964\\
    32&295&6765&14421\\
    36&1067&21164&52237\\
    40&3157&56823&154680\\
    44&8073&135728&395560\\
    48&18465&295545&904761\\
    52&38685&596980&1895536\\
    56&75500&1133187&3699499\\
    60&138986&2041480&6810300\\
    64&243611&3517605&11936925\\
    68&409544&5832828&20067614\\
    72&664191&9354095&32545333\\
    76&1043996&14567520&51155776\\
    80&1596508&22105457&78228865\\

\hline
$N$&$I(8,4,N,3)$&$[q^N]\nu_{E_{8}}(3;q)$&$S(8,1,N,3)$\\
%\hline
    
    %24&1&1&0\\
    %28&1&36&19\\
    32&8&331&378\\
    36&61&1752&2964\\
    40&295&6765&14421\\
    44&1067&21164&52237\\
    48&3157&56823&154680\\
    52&8073&135728&395560\\
    56&18465&295545&904761\\
    60&38685&596980&1895536\\
    64&75500&1133187&3699499\\
    68&138986&2041480&6810300\\
    72&243611&3517605&11936925\\
    76&409544&5832828&20067614\\
    80&664191&9354095&32545333\\
    84&1043996&14567520&51155776\\
    88&1596508&22105457&78228865\\
\hline
$N$&$I(8,4,N,4)$&$[q^N]\nu_{E_{8}}(4;q)$&$S(8,1,N,4)$\\
%\hline

    %32&1&1&0\\
    %36&1&36&19\\
    40&8&331&378\\
    44&61&1752&2964\\
    48&295&6765&14421\\
    52&1067&21164&52237\\
    56&3157&56823&154680\\
    60&8073&135728&395560\\
    64&18465&295545&904761\\
    68&38685&596980&1895536\\
    72&75500&1133187&3699499\\
    76&138986&2041480&6810300\\
    80&243611&3517605&11936925\\
    84&409544&5832828&20067614\\
    88&664191&9354095&32545333\\
    92&1043996&14567520&51155776\\
    96&1596508&22105457&78228865\\
\hline
\end{tabular}
}
\end{table}
\begin{table}[ht]
\centering \caption{\label{table4} Bounds on $A(n,d,N,r)$ with $L=BW_{16}$ and $r=2,3,4$}

\tiny
{
\begin{tabular}{|cccc|}
\hline
$N$&$I(16,4,N,2)$&$[q^N]\nu_{BW_{16}}(2;q)$&$S(16,1,N,2)$\\
%\hline  

    %32&1&1&0\\
    36&1&16&117\\
    40&82&306&14858\\
    44&2890&3984&526783\\
    48&49949&39235&9107278\\
    52&539795&310176&98422520\\
    56&4178302&2016996&761843656\\
    60&25184088&11005344&4591898687\\
    64&124915457&51463749&22776251653\\
    68&529944363&210557360&96626522164\\
    72&1977679995&767796630&360596985630\\
    76&6630474804&2535136560&1208956572561\\
    80&20297778673&7680579975&3700961644542\\
    84&57467324395&21588192576&10478208814512\\
    88&152025004051&56814408136&27719225738485\\
    92&378928483749&141077361984&69091293536850\\
    96&896068510238&332674600329&163383158366718\\        
    
\hline
$N$&$I(16,4,N,3)$&$[q^N]\nu_{BW_{16}}(3;q)$&$S(16,1,N,3)$\\
%\hline

    %48&1&1&0\\
    52&1&16&117\\
    56&82&306&14858\\
    60&2890&3984&526783\\
    64&49949&39235&9107278\\
    68&539795&310176&98422520\\
    72&4178302&2016996&761843656\\
    76&25184088&11005344&4591898687\\
    80&124915457&51463749&22776251653\\
    84&529944363&210557360&96626522164\\
    88&1977679995&767796630&360596985630\\
    92&6630474804&2535136560&1208956572561\\
    96&20297778673&7680579975&3700961644542\\
    100&57467324395&21588192576&10478208814512\\
    104&152025004051&56814408136&27719225738485\\
    108&378928483749&141077361984&69091293536850\\
    112&896068510238&332674600329&163383158366718\\
    
\hline
$N$&$I(16,4,N,4)$&$[q^N]\nu_{BW_{16}}(4;q)$&$S(16,1,N,4)$\\
%\hline

    %64&1&1&0\\
    68&1&16&117\\
    72&82&306&14858\\
    76&2890&3984&526783\\
    80&49949&39235&9107278\\
    84&539795&310176&98422520\\
    88&4178302&2016996&761843656\\
    92&25184088&11005344&4591898687\\
    96&124915457&51463749&22776251653\\
    100&529944363&210557360&96626522164\\
    104&1977679995&767796630&360596985630\\
    108&6630474804&2535136560&1208956572561\\
    112&20297778673&7680579975&3700961644542\\
    116&57467324395&21588192576&10478208814512\\
    120&152025004051&56814408136&27719225738485\\
    124&378928483749&141077361984&69091293536850\\
    128&896068510238&332674600329&163383158366718\\

\hline
\end{tabular}
}
%\end{multicols}
\end{table}

%%%%%%%%%%%%%%%%%%%%%%%%%%%%%%%%%%%%%%%%%%%%%%%%%%%%
\section{Asymptotic bounds on $A(n,d,N,r)$}\label{five}
We assume that $r$ is fixed,  that $N \rightarrow \infty$, and that $n \sim \eta  N/r,\,d\sim \delta N$ 
for some constants $\eta,\delta$ with $ \eta \in (0,1),$ and $\delta \ge 0.$ Because each codeword has weight $N,$ the triangle inequality in the Manhattan metric shows that $\delta \in (0,2).$ 
Denote by $R$ the asymptotic exponent of $A(n,d,N,r)$, that is
$$R\defeq \limsup \frac{1}{N}\log_2 A(n,d,N,r). $$
The asymptotic form of Theorem \ref{johnson} shows that $\delta \in (0,1)$ whenever $R\neq 0.$

Let 
$$L(x)=x\log_2 x+ \log_2 (x+\sqrt{x^2+1})-x \log_2(\sqrt{x^2+1}-1).$$
It was proved in \cite{GS} that when $x \rightarrow \infty$ and $ e\sim \epsilon n$ 
$$ \lim\frac{1}{n}\log_2 V(n,e) =L(\epsilon). $$
For convenience, let $$H(q)\defeq -q\log_2 q-(1-q)\log_2 (1-q)$$ denote the binary entropy function and let
$$f(x,y,z)\defeq [1-y+y/x]H(\frac{y}{y+x(1-y)})-(y/x)L(\frac{xz}{y}).$$
We establish the asymptotic version of Theorem~\ref{fonda}.
{\thm \label{asymp} With the above notation we have
$$ f(r,\eta,\delta)\le R\le f(r,\eta,\delta/2).$$
 }
\begin{proof}
%Combine Lemma \ref{2} and Lemma \ref{3} with the standard arguments.
The result follows from Theorem \ref{fonda} by standard entropic estimates for binomial coefficients  for the numerator and 
the result on large alphabet Lee balls from \cite{GS} for the denominators.
\end{proof}

\begin{figure}[ht]
	\centering{
		\includegraphics[width=8.5cm]{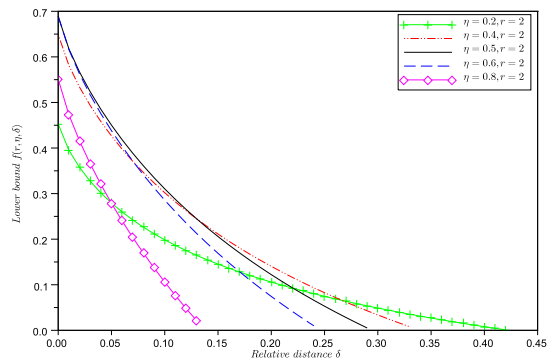}}
	\caption{\label{fig:r21} Graphs of $f(r,\eta,\delta)$ for $r=2$ and $\eta=0.2,0.4,0.5,0.6,0.8$ }
\end{figure}

\begin{figure}[ht]
	\centering{
		\includegraphics[width=8.5cm]{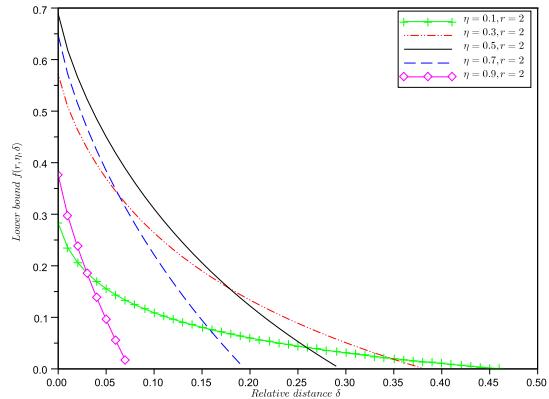}}
\caption{	\label{fig:r22}Graphs of $f(r,\eta,\delta)$ for $r=2$ and $\eta=0.1,0.3,0.5,0.7,0.9$ }
\end{figure}

In Fig. \ref{fig:r21} and \ref{fig:r22}, the graphs of the asymptotic lower bound curve $f(r,\eta,\delta)$ with different parameters $\eta$ and $r=2$ show that the rate $R$ is higher when $\eta$ is around $0.5$. 
%For all $\eta \in (0,1)$, $f(1,\eta,\delta)$ stays higher than other curves with $r\geq 2$. For $\eta$ close to the upper bound and $r$ large, $f(r,\eta,\delta)$ stays higher in a small interval of $\delta$.

%\setlength{\columnsep}{0.05cm}
%\begin{multicols}{2}

%\begin{figure}[ht]
%	\centering{
%		\includegraphics[width=8.5cm]{r41.pdf}}
%	\caption{\label{fig:r41} Graphs of $f(r,\eta,\delta)$ for $r=4$ and $\eta=0.2,0.4,0.5,0.6,0.8$ }
%\end{figure}

%\pagebreak
%\begin{figure}[ht]
%	\centering{
%		\includegraphics[width=8.5cm]{r42.pdf}}
%\caption{	\label{fig:r42}Graphs of $f(r,\eta,\delta)$ for $r=4$ and $\eta=0.1,0.3,0.5,0.7,0.9$ }
%\end{figure}

%\begin{figure}[ht]
%	\centering{
%	\includegraphics[width=8.5cm]{eta05.pdf}}
%\caption{	\label{fig:eta05}Graphs of $f(r,\eta,\delta)$ for $\eta=0.5$ and $r=2,3,4,5,6$ }
%\end{figure}

%\end{multicols}

%\pagebreak

\section{Conclusion and open problems}
\label{six}
We approached a problem of binary coding for the Levenshtein distance by using lattices for the Manhattan metric.
These lattices are obtained by Construction A applied to binary and quaternary codes. Since decoding these lattices for the Manhattan metric can be reduced to decoding the constructing code
for the Lee distance \cite{CJC}, it is worth to investigate the decoding of $\Z_4-$ codes beyond the Klemm's code considered here. Another approach would be to consider $\Z_4-$codes with a known decoding algorithm ({\it{e.g.}}, Preparata \cite{HK+}, Goethals \cite{HK}, Calderbank-MacGuire \cite{R}) and look at the performance of the corresponding lattices.

More generally, it is worth considering larger alphabets like $\Z_8,\Z_{16},$ when building lattices
in higher dimensions. The Lee decoding problem for such codes is completely open.
Moving away from Construction A, finding the densest lattice for the Manhattan metric in a given dimension is still a deep and fundamental open problem. 

Finally, turning to the deletion channel, what allowed us to use algebraic coding techniques was our working hypothesis; the runlengths of each codeword is larger than $r$, the maximum number of deletions that can occur over the transmission period. Extending these techniques to the case where the working hypothesis does not necessarily hold is an important and challenging open problem.
\section{Acknowledgments}
The authors would like to thank Jean-Claude Belfiore for helpful discussions.

%%%%%%%%%%%%%%%%%%%%%%%%%%%%%%%%%%

\end{document}